\documentclass[twocolumn]{aastex631}

\shorttitle{Radio polarization of millisecond pulsars}
\shortauthors{Sur et al.}
\usepackage{amsmath}
\graphicspath{{./}{figures/}}

\usepackage{tipa}
\usepackage{upgreek}
\usepackage{hyperref}

\begin{document}

\title{Radio polarization of millisecond pulsars with multipolar magnetic fields}

\author[0000-0001-6635-5080]{Ankan Sur}
\affiliation{Center for Computational Astrophysics, Flatiron Institute, Simons Foundation, New York, NY 10010, USA}
\affiliation{Department of Astrophysical Sciences, Princeton University, 4 Ivy Ln, Princeton, NJ 08544, USA}
\affiliation{Nicolaus Copernicus Astronomical Center, Polish Academy of Sciences, Bartycka 18, 00-716, Warsaw, Poland
}

\author[0000-0002-0108-4774]{Yajie Yuan}
\affiliation{Center for Computational Astrophysics, Flatiron Institute, Simons Foundation, New York, NY 10010, USA}
\affiliation{Department of Physics, Washington University in St. Louis, One Brookings Drive, St. Louis, MO 63130, USA}

\author[0000-0001-7801-0362]{Alexander Philippov}
\affiliation{Center for Computational Astrophysics, Flatiron Institute, Simons Foundation, New York, NY 10010, USA}
\affiliation{Department of Physics, University of Maryland, College Park, MD 20742}



\begin{abstract}

NICER has observed a few millisecond pulsars where the geometry of the X-ray emitting hotspots on the neutron star is analyzed in order to constrain the mass and radius from X-ray light curve modeling. One example, PSR J0030+0451, is shown to possibly have significant multipolar magnetic fields at the stellar surface. Using force-free simulations of the magnetosphere structure, it has been shown that the radio, X-ray, and gamma-ray light curves can be modeled simultaneously with appropriate field configuration. An even more stringent test is to compare predictions of the force-free magnetosphere model with observations of the radio polarization. This paper attempts to reproduce the radio polarization of PSR J0030+0451 using a force-free magnetospheric solution. As a result of our modeling, we can reproduce certain features of the polarization well.

\end{abstract}

\keywords{Pulsars (251) --- radio astronomy(1736)}


\section{Introduction} \label{sec:intro}

{A millisecond pulsar (MSP) has an extremely small spin period ($P\sim1-30$ ms) and differs from ordinary pulsars in that it has an extremely small spindown rate and exists in a binary system \citep{Lorimer2008}. Their short periods are thought to be caused by accretion of matter from a donor star \citep{Alpar1982}. A pulsar's magnetic pole emits cones of bright radio emission when it rotates rapidly, sweeping around like a lighthouse, and the interval between pulses is as precise as that of an atomic clock.} Radiation emitted by pulsars is generally believed to be produced in pair plasma outflow streaming along the dipolar magnetic field lines \citep{Philippov2020, Melrose2021, Philippov2022}. However, it is still theoretically unclear how pulsars produce such coherent sources of radio waves. Nevertheless, over the past decades, pulsars have proven to be fascinating laboratories for studying fundamental physics. For example: (1) measuring their masses and radii can constrain the equation of state for nuclear matter \citep{Raajimakers2019,Bogdanov2021}, (2) measuring their motion can help test general relativity (GR) \citep{Kramer2006,Kramer2009}, (3) Pulsar timing can be used to detect gravitational-wave background \citep{Nano2018,Nano2020}, and finally (4) they are good probes for studying the local environments and properties of their host galaxies \citep{Coles2015,Jones2017}. 

We can get constraints on the emission physics from the pulsar magnetosphere through multiwavelength analysis. For example, NICER has obtained detailed X-ray observations of hot spots present in the pulsar PSR J0030+0451 \citep{Miller2019, Riley2019} which has allowed us to constrain its magnetic field geometry \citep{Bilous2019}. Their results showed that the magnetic field is far from a simple dipole but favors multipolar components at the stellar surface. The shape and location {of the hot spots observed in thermal X-rays} are the footprints of open magnetic field lines. {Here, active pair-production occurs, which leads to particle bombardment of the stellar surface and production of X-rays}. {The thermal X-ray lightcurves of PSR J0030+0451, and, in addition, radio and $\gamma$-rays, produced in the outer magnetosphere}, were successfully modeled by \cite{Chen2020, Kalapotharakos2021}. {{This modeling}} confirmed that a multipolar magnetic field is key in {{understanding}} this system and can be further used to constrain the magnetic inclination angle.

The magnetic field in ordinary pulsars has been shown to evolve through the Hall effect and Ohmic dissipation in the crust mediated by free electrons \citep{Cumming2004, Pons2007, Gourgouliatos2014, Bransgrove2018}, and {ambipolar diffusion driven by binary scattering processes in electron-proton-neutron plasma} \citep{Goldreich1992,Castillo2017,Bransgrove2018}. Modeling the interior and exterior field consistently is difficult, and in fact, most models for the interior field in the star do not model the magnetosphere and just have some fixed exterior boundary, and vice versa. The dynamical interplay between these regions is crucial to understand some magnetospheric phenomenology. Global magnetohydrodynamics simulations with initial dipolar magnetic field for ordinary pulsars have shown that after birth the system evolves to a state with significant power in the multipolar components \citep[e.g.,][]{Sur2020}. 
{Higher-order multipoles near the surface of the star have been proposed to activate pair production that is supposed to power the radio emission \citep{Gil2006}. However, these multipoles would dissipate within a few million years due to Ohmic decay \citep{Gourgouliatos2014} even if they were generated after an ordinary pulsar is recycled into a MSP.}
The formation process of MSPs through accretion may also change the magnetic field configuration of MSPs, either through burial of magnetic field \citep{1990Natur.347..741R,2001PASA...18..421M,2004MNRAS.351..569P}, or due to field migration as the neutron star spins up \citep{1991ApJ...366..261R,1993ApJ...408..179C,1998ApJ...493..397C}. The magnetic field is therefore challenging to model in MSPs. The dipolar field model has been commonly used in various studies, such as determining the magnetic field strength from the dipole spindown and explaining the inverse relationship between spin period and pulse width. However, the presence of non-dipolar magnetic field configurations has an impact on various aspects of pulsar astrophysics, {including birth velocities \citep{Radhakrishnan1984,Bailes1989}} and the interpretation of multi-wavelength magnetospheric emission. Therefore, it is crucial to have a precise representation of the magnetic field when studying MSP-related phenomena.


\begin{figure}
\centering
\includegraphics[width=8.5cm]{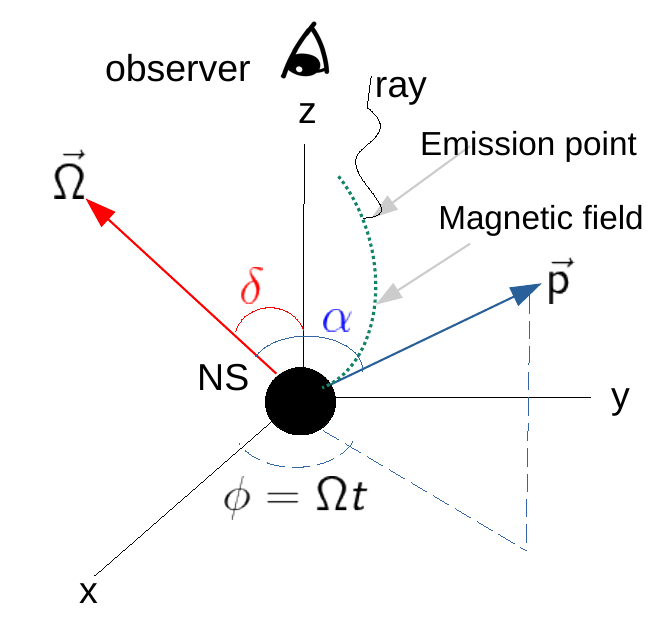}
\caption{The rotating NS is located in the center of XYZ frame with its magnetic moment axis ($\vec{p}$) and rotation axis ($\vec{\Omega}$). $\vec{\Omega}$ lies in the x-z plane. The observer is located on the z-axis, the green dotted line shows the magnetic field line, while the curly black solid line shows a ray propagating towards the observer.}
\label{fig1}
\end{figure}

One of the most important tools to connect the pulsar magnetic field geometry to observations is the radio light curve and polarization modeling. In the past, the rotating vector model (RVM) \citep{Radhakrishnan1969} was widely used in this respect. The model assumes that the emission comes from near the magnetic pole, and the polarization is determined by the direction of the magnetic field at the emission point. As the line of sight cuts across the magnetic pole, the plane of linear polarization sweeps a characteristic S-shaped swing over a single pulse period. Radio pulsar mean profiles are typically interpreted according to the hollow cone model \citep{Manchester1977}. The main assumptions in this model are the following: (1) the emission is generated in the inner magnetosphere (where the magnetic field ($\mathbf{B}$) is thought of as a dipole), (2) the emission travels in a straight line, (3) cyclotron absorption may be ignored, and finally (4) the polarization is determined at the emission point. Analytically, the change of the position angle (PA) with respect to the rotation phase ($\phi$) of the pulsar is given by
\begin{align}
    \rm PA &= \rm arctan(B_y/B_x) \\ 
     &=\rm arctan \bigg( \frac{\sin\alpha{\,\sin\phi}}{\sin\alpha\cos\delta\cos\phi - \sin\delta\cos\alpha}\bigg)
    \label{eq1}
\end{align}
where $\alpha$ is the angle between the magnetic moment axis and the rotation axis while $\delta$ is the angle between the rotation axis and the line of sight of the observer. According to the RVM, PA is determined solely by the projection of magnetic field on the sky plane (see figure \ref{fig1}). In this convention, the x-axis is along the projection of the angular velocity $\boldsymbol{\Omega}$ on the sky plane.

{Another modification to the RVM was the model presented by \cite{BCW1991} (BCW) which incorporates relativistic effects, such as aberration due to significant corotation component of the plasma velocity and retardation. The BCW model calculates the position angle by finding the direction of the acceleration of radiating charged particles at the emission point. The relativistic effect causes the polarization profile to lag the intensity profile. The magnitude of the lag of the inflection point of the PA profile in the presence of aberrations and retardations can be calculated as follows:
\begin{equation}
    \Delta_{\phi} = \frac{8\pi r_{\rm em}}{c P}
    \label{bcw_delay}
\end{equation}
where $P$ is the period, $r_{\rm em}$ is the emission radius, and $c$ is the speed of light. Based on the observed lag, the  emission radius can be inferred.} 

Although the RVM with a dipolar field successfully explains pulse profiles of many pulsars qualitatively \citep[e.g.,][]{Manchester1975, Johnston2023} and even quantitatively \citep[e.g.,][]{Desvignes2019}, observations of MSPs often show flat, distorted, or even random PA profiles hinting towards possible non-dipolar configurations \citep{Backer1976}. In addition, propagation of radio waves through magnetospheric pair plasma can further change the polarization properties  \citep{Petrova2000,Wang2010,Beskin2012,Hakobyan2017,Alisa2020} which are not included in the RVM model. {It is important to note that some MSPs have flux densities similar to ordinary pulsars and their radio profiles are only marginally more complex \citep{Kramer1998}. Nonetheless, physical conditions in magnetospheres of MSPs and their surface magnetic field structure could differ considerably owing to its evolutionary history, which can result in changes to their observed appearance \citep{Philippov2022}.}

In this paper, we are interested in comparing radio polarization with observations for the MSP PSR J0030+0451 taking into account a multipolar magnetic field configuration {as obtained in \cite{Chen2020}}. 
Developing a better understanding of the multipolar structure and surface field will help to constrain not only radio emission sites, but also the evolution of the magnetic field. {As a first step, we use a modified RVM: we assume that the radio emission is produced as plasma normal modes at a radius $r_{em}$ which then propagates along a straight line, where its PA evolves adiabatically following the magnetic field. The PA then freezes at a distance $h$ from the emission point where the magnetospheric plasma density has dropped so that the radio waves follow vacuum propagation afterwards. We take into account the aberration effect self-consistently, but neglect other propagation effects for now. With this approach, we systematically obtain PA sky maps and curves with $r_{em}$ and $h$ as free parameters, which we then compare with the observed PA signatures.} The paper is organized as follows: in section \ref{sec:model} we discuss the magnetic field model, in section \ref{sec:aberration} we derive an expression for PA considering aberration, in section \ref{sec:method} we describe our method, in section \ref{sec:results} we present the results, and finally, in section \ref{sec:conclusions} we discuss conclusions.

\begin{figure*}
    \centering
    \includegraphics[scale=0.3]{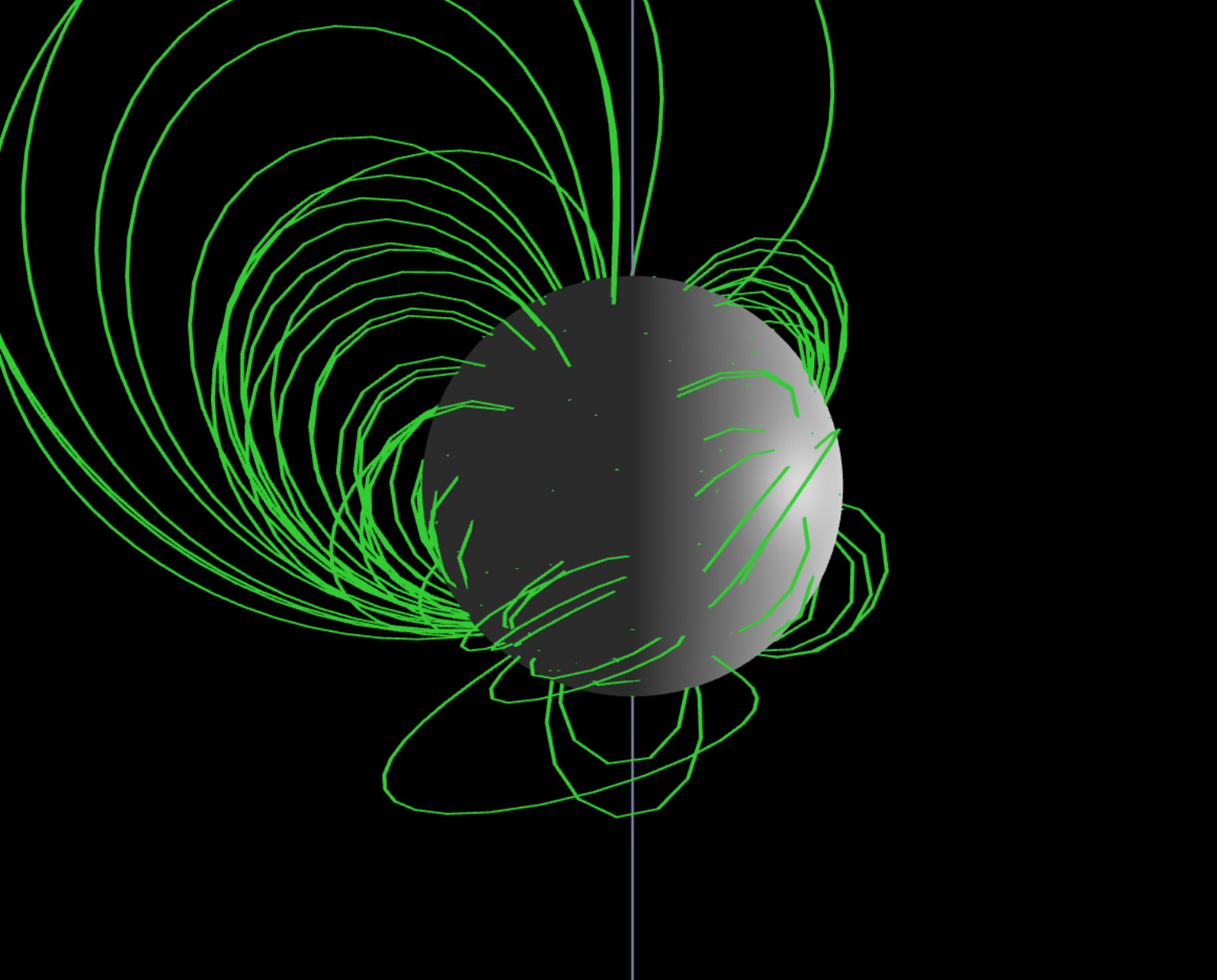}
    \includegraphics[scale=0.335]{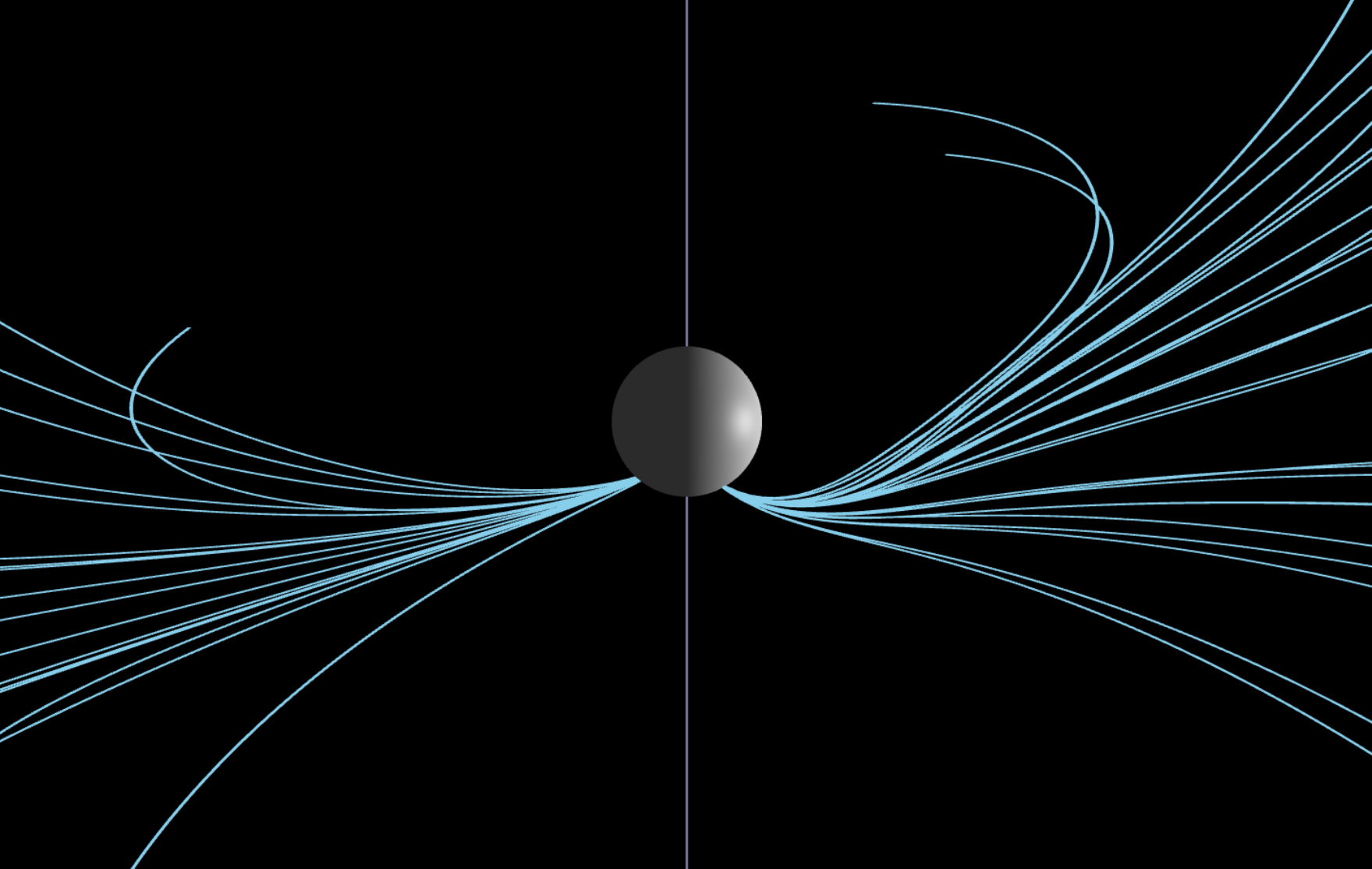}
    \caption{The vacuum magnetic field configuration as seen from the meridional view along the x-axis. The left figure shows the closed magnetic field lines with some of the quadrupolar loops while the right figure shows the open dipolar field lines. The rotation axis is along the z-axis and passes through the center of the star.}
    \label{fig:vacuum}
\end{figure*}

\begin{figure*}
    \centering
    \includegraphics[scale=0.32]{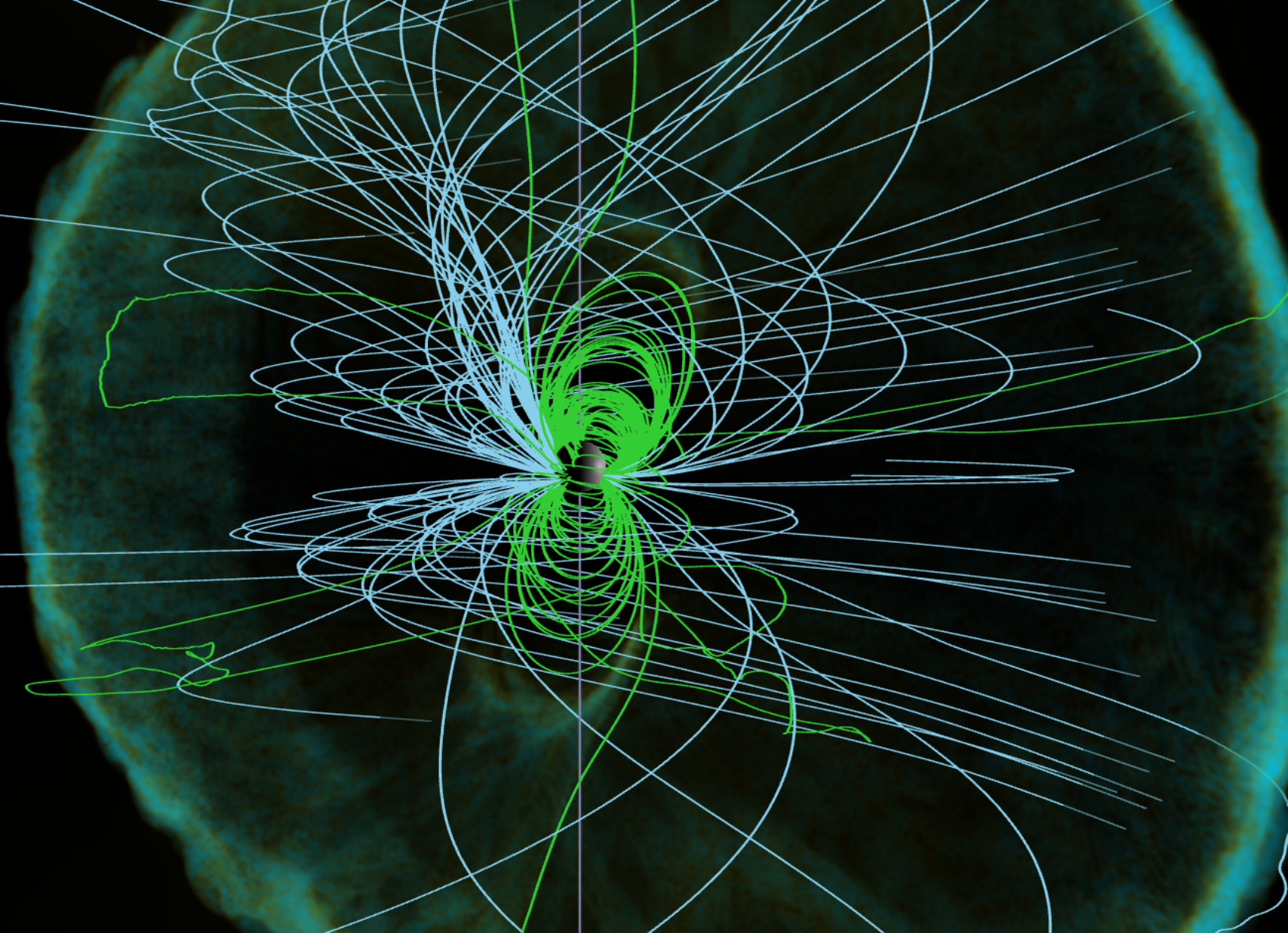}
    \includegraphics[scale=0.319]{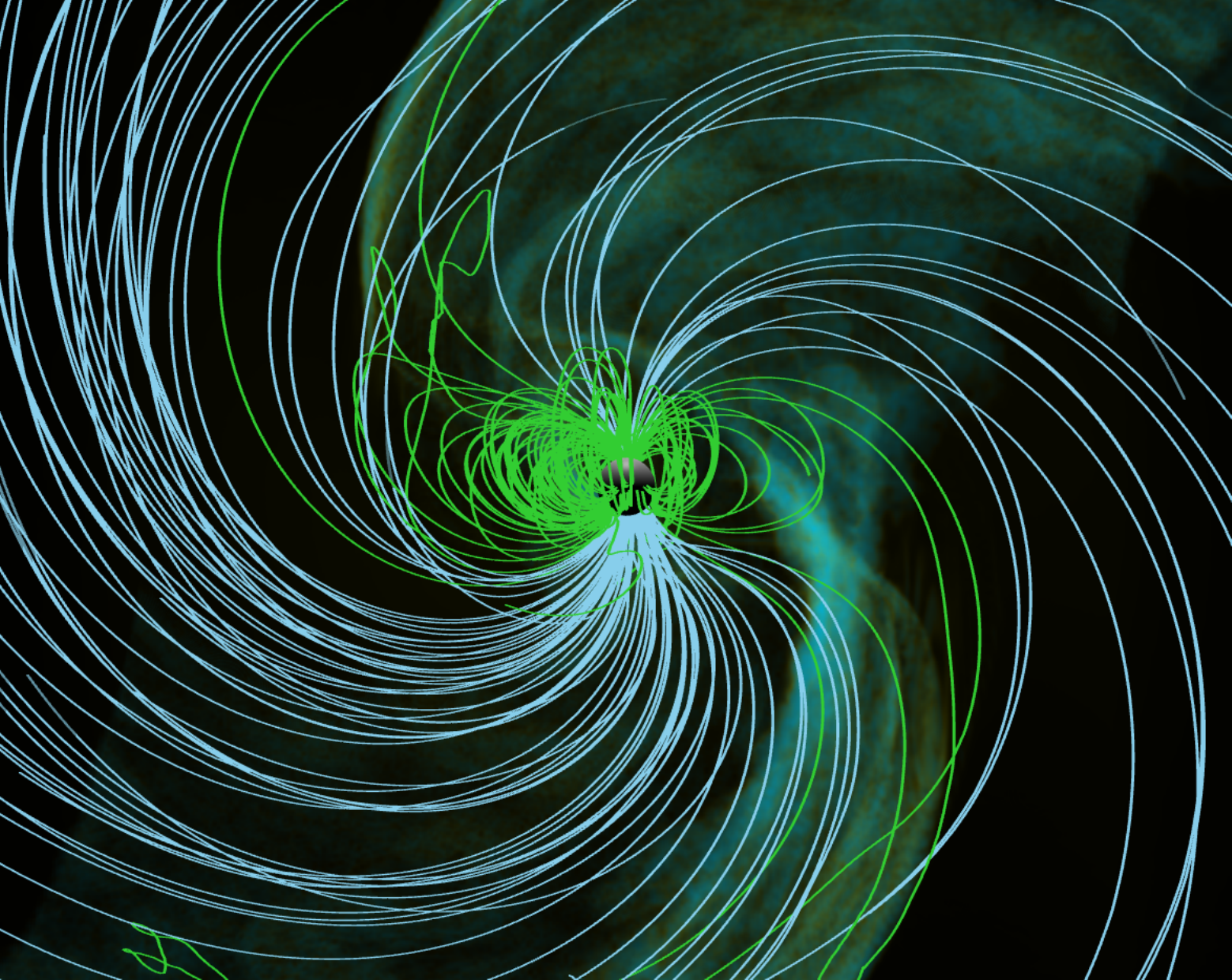}
    \caption{The force-free magnetic field configuration as seen from (left) meridional view (along x-axis), and (right) equatorial view. The green lines represent the closed field lines while the cyan lines represent the open field lines. Also shown in the background is the location of current sheets. {This field configuration was obtained in \cite{Chen2020}}.}
    \label{fig:Forcefree}
\end{figure*}

\section{The Magnetospheric Model} \label{sec:model}
At the polar caps of a plasma-filled magnetosphere, electric current flows on open field lines. It has been reported that PSR J0030+0451 has a spin period of 4.18 ms with hotspots that are beyond a simple dipole, requiring multipolar magnetic field components \citep{Bilous2019}. {We use the field configuration first deduced by} \cite{Chen2020}, who showed that a quadru-dipolar magnetic field can be used to reproduce the light curves in X-rays, gamma rays, and radio waves for PSR J0030+0451. The analytical form of this field in vacuum is given by
\begin{equation}
\mathbf{B} = \mathbf{B_d} + \mathbf{B_q}
\end{equation}
where the dipolar component is
\begin{equation}
    \mathbf{B_d} = -\frac{\mathbf{p}}{{r}^3} + \frac{\left(3\mathbf{p}\cdot\mathbf{r}\right)\mathbf{r}}{{r}^5}
\end{equation}
with the dipole moment $\textbf{p} = p_0(0,0.985,0.174)$, while the quadrupolar component is
\begin{equation}
    \mathbf{B_q} = -\frac{2}{r^5}\mathbf{Qr} + \frac{5}{{r}^7}(\mathbf{rQr^T})\mathbf{r}
\end{equation}
with
\begin{equation}
    \mathbf{Q} = p_0{R^2} \begin{pmatrix}
0.6 & 0 & 0\\
0 & -0.8 & -2 \\
0 & -2 & -2 
\end{pmatrix},
\end{equation}
{where $R$ is the neutron star radius.}
The dipole inclination angle is $80^{\circ}$ and the quadrupolar component is centered at $(0,0,-0.4R)$. {These exact set of parameters were used in \cite{Chen2020}}. A 3-dimensional view of the field is shown in figure \ref{fig:vacuum}. The field is symmetric about the y-z plane, and the polar caps are located in the southern hemisphere, one of them circular and the other crescent-shaped. 

The {{structure of the force-free magnetic field}} is shown in figure \ref{fig:Forcefree}. Beyond the light cylinder radius ($R_{\rm LC}$), which is located at $20R$ for PSR J0030, the field lines are opened, and opposite polarities are separated by a current sheet \citep{Spitkovsky2006,Kalapotharakos2009,Chen2014,Philippov2018,Hayk2019}.

\section{Effect of Aberration} \label{sec:aberration}
In addition to our force-free magnetic field solution, we assume that the outflowing plasma has $\upgamma = 1/\sqrt{1-v^2/c^2} \gg 1$ {moving along the magnetic field lines}. Let us also assume that the emission comes from a sphere of constant radius ($r_{em}$). 
{\footnote{We use this crude assumption as a demonstration of principle. Typically $r_{em}$ depends on plasma parameters, e.g. pair density. This is not well constrained and often assumed to have simple profiles, e.g. spherically symmetric distribution \citep{Petri2019}.}}
In the case of emission from one magnetic pole, there is a single point on this sphere where particles beam along {the observer direction} $\hat{\mathbf{n}}$  at a particular time. In ordinary pulsars, the emission direction is solely dependent on the direction of the magnetic field. But since PSR J0030+0451 is an MSP with spin period of 4.18 ms, the aberration angle at the emission point is approximately $\Omega r_{em}/c$ ($\sim {0.3}$ at $r_{em} = 5R$) which cannot be neglected compared to the angular size of the emission cone ($1/\upgamma$)\footnote{Pair production discharge in MSPs is not well understood and as a result this quantity may not be small.}. {As inferred from the aberration-retardation modeling, typical values of the emission radius for MSPs are less than 100 km, which translates to $\sim 0.05-0.5 \, R_{LC}$ \citep{Rankin2017}. On the other hand, ordinary pulsars emit radiation at a height $\sim 300-1000$ km \citep{Gupta2003, Dyks2004, Johnston2023} where the magnetic field is dipolar. The aberration angle of an ordinary pulsar, for example, PSR J1808-0813 with period $P=0.876$ s, is $\Omega r_{em}/c \sim 1\times10^{-7}$, where $r_{em}$ is taken to be $\sim 693$ km \citep{Mitra2023}. This is negligibly small compared to the aberration angle of PSR J0030+0451. In this work, we assume $0.1 \,R_{LC}\leq \,r_{\rm em} \leq 0.5 \,R_{LC}$.}

{{The location of the emission point, $\mathbf{r}_{\rm em}$, is determined by demanding the direction of the particle propagation, along which the radiation is beamed, to be along the observed line of sight, $\hat{\mathbf{n}}$  \citep{Wang2010,Beskin2012}. The speed of the outflowing plasma is {close} to the speed of light, $c$, resulting in the following condition}}
\begin{equation} \label{eq:prop_direction}
    \hat{\mathbf{n}} = \frac{\boldsymbol{\Omega} \times \mathbf{r}}{c} + \kappa \mathbf{b},
\end{equation}
{where $\mathbf{b}=\mathbf{B}/B$ is a unit vector in the direction of the magnetic field, and $\kappa$ is a dimensionless constant. Its magnitude can be obtained by demanding that the flow is away from the star, and $\hat{\mathbf{n}}\cdot\hat{\mathbf{n}} = 1$.} 

\subsection{Normal modes}
{Normal modes in plasma physics refer to the oscillations and pattern of waves that a plasma can sustain due to interaction between the electromagnetic forces and charged particles. Assuming a cold pair plasma, the wave dispersion relation in the rest frame is given by \citep{Philippov2022}:
\begin{equation}
    (\omega^2 - c^2 k^2)\big[(\omega^2 - c^2 k_{||}^2)(1 - \omega_p^2/\omega^2) - c^2/k_{\perp}^2\big] = 0
    \label{eq:dispersion}
\end{equation}
where $\omega_p$ is the plasma frequency, $\omega$ is the wave frequency, and $k_{||}$ and $k_{\perp}$ 
{are the components of the wave vector $\mathbf{k}$ parallel and perpendicular to the background magnetic field, respectively.}
The solution to equation \ref{eq:dispersion} gives three linearly polarized waves: the extraordinary mode (X-mode) having the electric field ($\mathbf{E}$) perpendicular to both the background magnetic field and $\mathbf{k}$-vector; and two ordinary modes (O-modes) with the $\mathbf{E}$-field in the plane of $\mathbf{k}$-vector and background magnetic field. In the pulsar magnetosphere, only radio waves corresponding to plasma normal modes can propagate and escape.}

\subsection{The new Position Angle}
Given that we found the location of the emission point, we next derive the polarization as seen by an observer on Earth. {Radio waves are emitted as plasma normal modes {as discussed before}. In the nearly force-free magnetosphere, the plasma is rotating together with the NS so the rotation will affect the plasma normal modes. Here we first obtain the normal modes in the corotating frame, then transform to the {laboratory} frame, because the normal modes are relatively easy to write down in the corotating frame. In the corotating frame, {the background electric field is zero and the polarization of the wave modes is determined by the background magnetic field and the wave vector in the same way as in the plasma rest frame} (see, e.g., \citealt{Wang2007}). Transforming back to the {laboratory} frame, we can obtain the polarization as seen by a distant observer.} In what follows, we consider the limit where the refractive index $n\equiv ck/\omega\approx1$. Here, the wave vector in the {laboratory} frame is \textbf{k} and the frequency is $\omega$. The {background} magnetic field in the {laboratory} frame is \textbf{B}, and the {magnetospheric} electric field is $\textbf{E} = -\boldsymbol{\beta\times\mathbf B}$, where $\boldsymbol{\beta = \Omega \times r}/c$ is the dimensionless rotation velocity. In the corotating frame (we use primed quantities for the corotating frame), {the background magnetic field becomes}
\begin{equation}
    \mathbf{B^{\prime}} = \frac{1}{\Gamma}\mathbf{B} + \left(1 - \frac{1}{\Gamma}\right)\frac{\mathbf{B}\cdot\boldsymbol{\beta}}{\beta^2}\boldsymbol{\beta},
\end{equation}
where $\Gamma=1/\sqrt{1-\beta^2}$ is the Lorentz factor corresponding to the rotation velocity. The wave frequency in the corotating frame is
\begin{equation}
    \omega^{\prime} = \Gamma\left(\omega - \mathbf{k}\cdot\boldsymbol{\beta}\right),
\end{equation}
and the wave vector becomes
\begin{equation}
    \mathbf{k^{\prime}} = \mathbf{k} + \frac{\left(\Gamma-1\right)}{{\beta^2}} (\mathbf{k}\cdot\boldsymbol{\beta})\boldsymbol{\beta} - \Gamma\omega\boldsymbol{\beta}.
\end{equation}
Considering an O mode with electric field $\mathbf{E}^{\prime}_w$ in the plane of the wave vector $\mathbf{k}'$ and the background magnetic field $\mathbf{B}'$, we can write the electric and magnetic field in the wave as the following:
\begin{equation}
    \mathbf{E}^{\prime}_w = \left(\mathbf{B^{\prime}} - \frac{\boldsymbol{k}^{\prime}\cdot\mathbf{B}^{\prime}}{k'^2}\mathbf{k}^{\prime}\right)E_0
    \label{eq13}
\end{equation}
\begin{equation}
    \mathbf{B}^{\prime}_w = \frac{1}{k^{\prime}}\left(\mathbf{k^{\prime}}\times\mathbf{B^{\prime}}\right)E_0
        \label{eq14}
\end{equation}
Now, {transforming} back to the {laboratory} frame, we obtain the wave electric field
\begin{multline}
   \mathbf{E}_w = \frac{1}{\beta^2}(\mathbf{E}^{\prime}_w\cdot\boldsymbol{\beta})\boldsymbol{\beta}+ \\
   \Gamma \left(\mathbf{E}^{\prime}_w - \frac{1}{\beta^2}(\mathbf{E}^{\prime}_w\cdot\boldsymbol{\beta})\boldsymbol{\beta}-\boldsymbol{\beta}\times\mathbf{B^{\prime}}_w\right)
   \label{eq15}
\end{multline}
{Plugging equations (\ref{eq13}) and (\ref{eq14}) in equation (\ref{eq15}), and after some algebra\footnote{This calculation agrees with the direct normal mode analysis performed in the laboratory frame in \cite{Petrova2000} and \cite{Beskin2012}.}, we get}
\begin{equation}
    \mathbf{E}_w = \frac{\omega^2E_0}{{k^{\prime}}^2}\bigg(\mathbf{B}_{\perp} + \boldsymbol{\beta}_{\perp}(\mathbf{B}\cdot\mathbf{\hat{k}}) - \mathbf{B}_{\perp}(\mathbf{\hat{k}}\cdot\boldsymbol{\beta}) \bigg)
    \label{eq18}
\end{equation}
where we have assumed $\omega=c k${, and $\perp$ indicates the component perpendicular to the wave vector $\mathbf{k}$}. Therefore, if we choose a coordinate system such that $\mathbf{k}$ is along $\hat{z}$, the wave polarization in the {laboratory} frame is given by
\begin{equation}
    \frac{E_x}{E_y} = \frac{B_x + \beta_xB_z - B_x\beta_z}{B_y + \beta_y B_z - B_y\beta_z}
\end{equation}
Thus, the new position angle including the effect of aberration is given by
\begin{equation}
    \rm PA = \arctan \left(\frac{B_y + \beta_y B_z - B_y\beta_z}{B_x + \beta_xB_z - B_x\beta_z} \right)
    \label{eq:PA_aber}
\end{equation}

\section{Method} \label{sec:method}
\begin{figure*}
    \centering
    \includegraphics[scale=0.465]{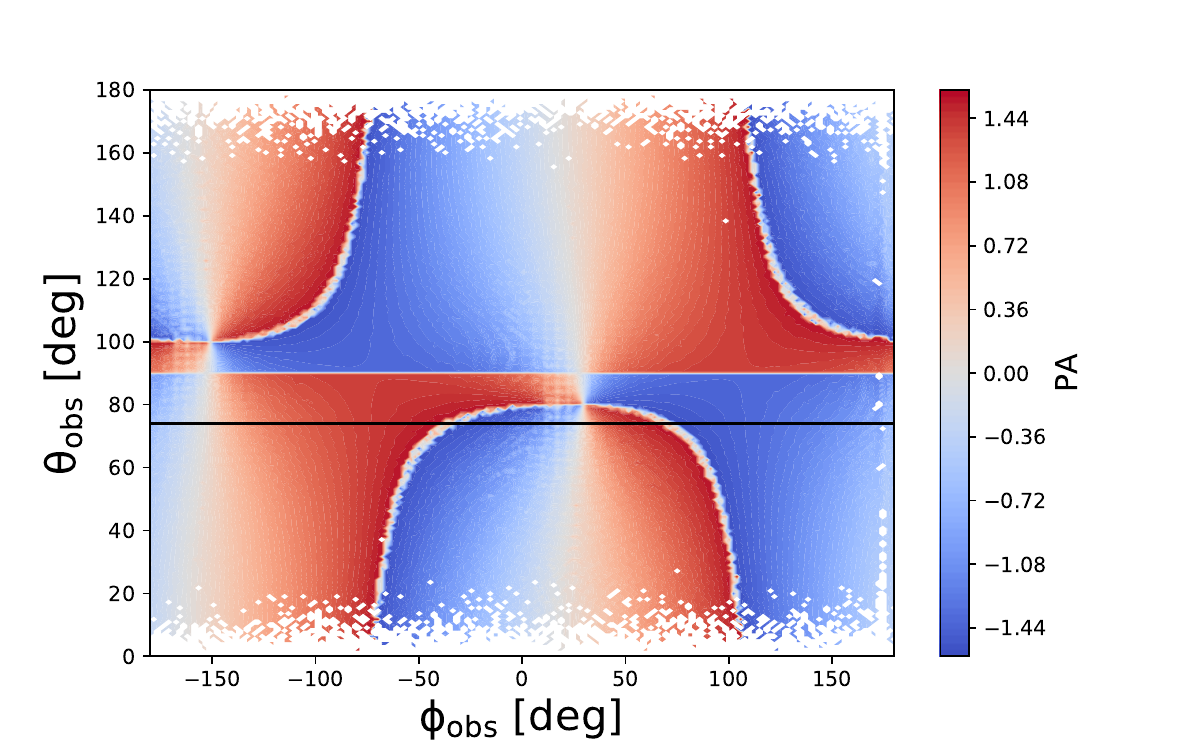}
    \includegraphics[scale=0.463]{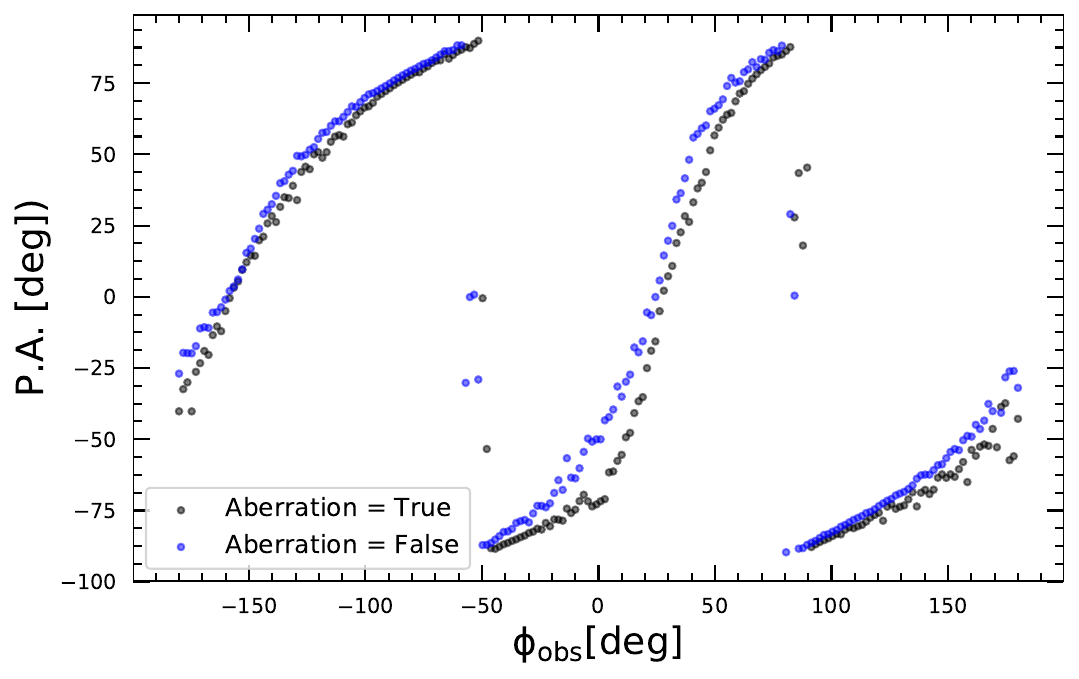}
    \caption{Left panel: the sky map for a vacuum dipolar magnetic field with an inclination angle of $\alpha=80^{\circ}$, and emission radius $r_{\rm em}=3R$. The color scale represents PA values while the horizontal black solid line represents PSR J0030+0451's viewing angle from earth. Right panel: the corresponding PA curves with and without aberration were observed from a viewing angle of 79 degrees and freezing height $h=0.1R$.}
    \label{fig:dip}
\end{figure*}

In order to obtain the PA curves, we first construct sky maps of the polarization. {Since the force-free solution reaches a steady state in the corotating frame, we just need one snapshot from the {force-free} simulation to create the all sky map. As a demonstration of principle, we consider emission coming from a spherical surface with radius $r_{\rm em}$. {This is the point in the pulsar magnetosphere where an outgoing radio wave is assumed to be emitted. 
Because of the large uncertainties in the emission physics, we consider a large range of emission heights, from close to the surface all the way up to 10 times the radius of the pulsar\footnote{Note that PSR J0030+0451 has $R_{LC}\approx 20 R$, where $R$ is its radius.}.}

Suppose each emission point has a polar angle $\theta$ and an azimuth angle $\phi$, the emission direction $\hat{\mathbf{n}}$ is determined using equation (\ref{eq:prop_direction}). We denote the polar angle of $\hat{\mathbf{n}}$ as $\theta_e$ and its azimuth angle as $\phi_e$. After emission, the ray propagates in a straight line, so it will reach an observer with polar angle $\theta_{\rm obs}=\theta_e$, and the corresponding phase is determined by \citep[e.g.,][]{Bai2010}
\begin{equation}
    \phi_{\rm obs} = -\phi_{\rm e} - \hat{\mathbf{n}}\cdot\mathbf{r}/R_{\rm LC}
\end{equation}
which takes into account the phase delay due to finite light travel times.
This step enables the mapping from the emission point $(\phi, \theta)$ to the observer plane $(\phi_{obs},\theta_{obs})$. For each ray, the polarization first evolves following the background magnetic field, then it freezes at a distance $h$ from the emission point. {This height is the distance above the $\mathbf{r}_{\rm em}$ where the radio wave decouples from the plasma and propagates freely before reaching the observer without changing its polarization}. Therefore, we determine the polarization angle at the freezing point $\mathbf{r} = \mathbf{r}_{\rm em} + h \hat{\mathbf{n}}$ using the local magnetic field, taking into account the rotation of the magnetosphere during the time interval it takes the ray to propagate through a distance $h$}. 
The polarization angle is then obtained through equation (\ref{eq:PA_aber}): for each ray we put the observer along z and the rotation axis of the NS in the x-z plane, then we find the transverse components of the magnetic field $B_x$, $B_y$ and the rotation velocity $\beta_x$, $\beta_y$ {at the freezing point}. Note that numerically, both the numerator and the denominator in equation (\ref{eq:PA_aber}) are zero at the emission point by definition. Therefore, a small but finite $h$ is needed even if we consider the usual RVM limit where the polarization is determined immediately at the emission point. 

For our sky map, we select a $200\times200$ grid uniformly spaced in $\theta_{\rm obs}$ and $\phi_{\rm obs}$. A ray originating from the emission surface at $(\phi^i,\theta^j)$ carries the information of the PA and maps to a particular grid point $(\phi_{\rm obs}^k,\theta_{\rm obs}^l)$ on the sky map. Once we have our sky map, we can fix the observing angle $\theta_{\rm obs}$, and plot the PA as a function of $\phi_{\rm obs}$. 

\begin{figure*}
    \centering
   \includegraphics[scale=0.34]{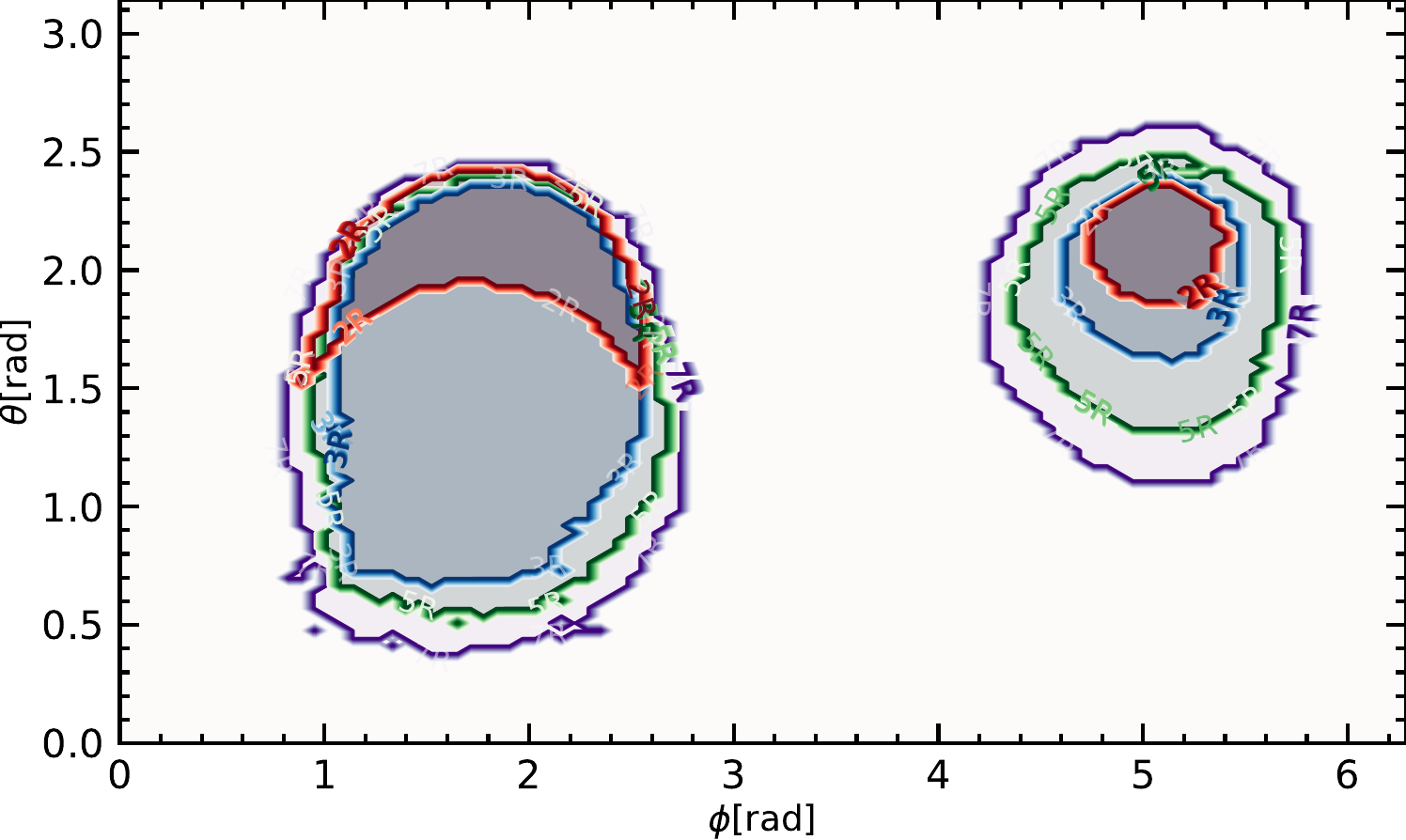}
    \includegraphics[scale=0.35]{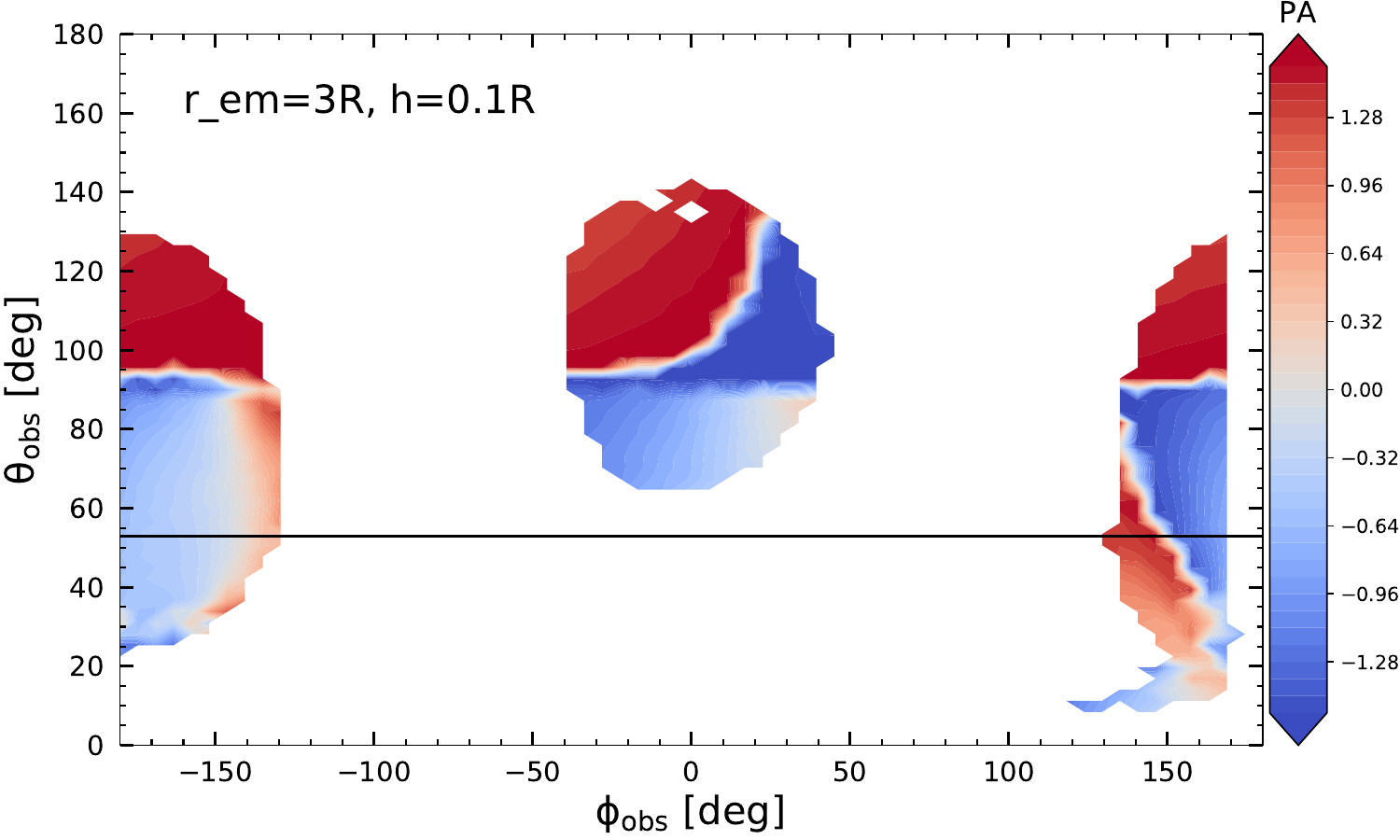}
    \caption{Left panel: the shape of the polar caps at $r=2R$ (red), $r=3R$ (blue), $r=5R$ (green), and $r=7R$ (violet), as seen by the observer. Close to the surface, one of the polar caps at 2R is crescent-shaped, while the other pole is more circular in nature. With increasing height, the poles become oval-shaped. Right panel: the sky map at $r_{em}=3R$ with the colorscale representing the PA values. Again, the horizontal black line is the line of sight of the observer set at 54 degrees.}
    \label{fig:PAgrid}
\end{figure*}

\section{Results} \label{sec:results}
\begin{figure*}
    \includegraphics[scale=0.355]{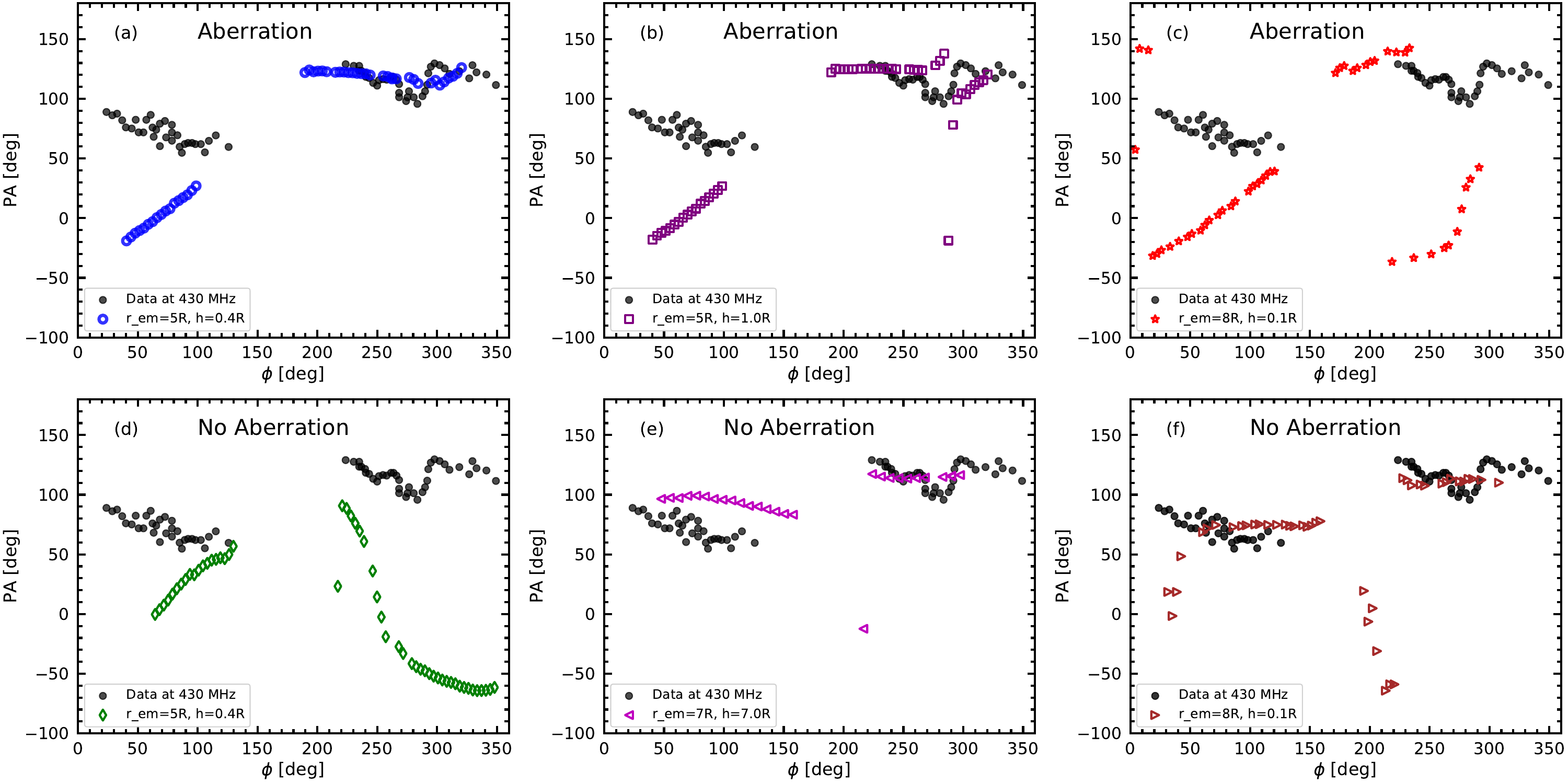}
    \caption{A comparison of the different PA curves represented by the scatter plots, obtained from our models by varying the freeing height (panel b), emission radius (panel c), and not including the effect of aberration (bottom-row) for the same parameters. The black dots shows the observation at 430 MHz.}
    \label{fig:PAcomp430}
\end{figure*}
As a first step, we perform a simple test to determine whether we get the usual result for a purely dipolar magnetic field. This is shown in figure \ref{fig:dip}, where two S-shaped swings from the two magnetic poles are clearly visible. The wider S curve (which gets broken into two parts by the phase beyond $180^{\circ}$) comes from the magnetic pole farther from the observer while the narrower one comes from the pole close to the observer. We also see the effect of aberration which makes the curves shift towards right, i.e. slightly towards higher $\phi_{\rm obs}$. {This effect is because aberration changes the orientation of the $\mathbf{E}$-vector in the wave and orientation of the polarization ellipse, which leads to changes in the PA profile. As per the RVM, most polarization observations are used to identify the 
{$\pmb{\Omega}-\pmb{\mu}$} meridional plane
which is the plane containing the magnetic axis $\pmb{\mu}$ and the pulsar rotation axis $\pmb{\Omega}$. In this scenario, the steepest gradient or the inflection point of the PA curve is contained within the {$\pmb{\Omega}-\pmb{\mu}$ plane}, and coincides with the midpoint of the pulse profile precisely. Nevertheless, in the observer's frame, as \cite{BCW1991,Dyks2004} pointed out, the aberration and retardation effects cause the PA inflexion point to appear delayed relative to the mid-point of the pulse profile. {The magnitude of this delay is given by:
\begin{equation}
    \Delta \phi_d \sim \mathcal{C} r_{\rm em}/ R_{LC}
\end{equation}
where $\mathcal{C}$ is some constant coefficient (see equation \ref{bcw_delay})}.

Next, we analyse the force-free magnetic field model for PSR J0030+0451. Due to the radio emission coming from the open field lines, we restrict our analysis to only using open field lines to obtain the sky map corresponding to the two magnetic poles. A view of the polar caps from different heights above the magnetic poles is shown in the left panel of figure \ref{fig:PAgrid}. As the height increases, the poles become more oval-shaped. The sky map at $r_{em}=3R$ is also shown in the right panel of figure \ref{fig:PAgrid} as an example.  

In figure \ref{fig:PAcomp430}, we plot the PA curves by fixing the viewing angle to $\theta_{\rm obs}=54^{\circ}$ as reported in \cite{Bilous2019}. Several different effects are explored, including the effect of varying the emission radius, varying the freezing height, and examining the effects of aberration. Our main aim is to reproduce the observed PA curves of PSR J0030+0451 at 430 MHz and at 1.4 GHz, as shown in \cite{Gentile2018}. When comparing different PA curves, we focus more on their shape than on their exact values. Let us first discuss the PA curve observed at 430 MHz which is shown in figure \ref{fig:PAcomp430}. It has two distinct parts corresponding to two different poles \citep{Gentile2018}. The one between $\phi_{\rm obs} = (20^{\circ} \text - 100^{\circ})$ comes from the smaller intensity pulse while the other comes from the higher intensity pulse. {The latter demonstrates a flat and distorted PA profile}, which hint that non-dipolar magnetic field effects are present. If {the emission comes from} very close to the surface, for example $r_{em} \sim 3R$, we can see only one of the magnetic poles (see figure \ref{fig:PAgrid}), and a very small part of the PA curve\footnote{If emission originates close to the surface, GR light bending and plasma refraction have to be taken into account and the rays will not propagate along straight lines. In this sense, our $r_{em}$ can be considered as the "effective" height where these effects freeze out.}. At very far from the surface ($r_{em}>8R)$, the PA curve becomes S-like due to the field being dipolar \footnote{The field is predominantly monopolar beyond the light cylinder, so it will be also interesting to see how this affects the PA curve.} there (see panel c in figure \ref{fig:PAcomp430}). Varying the freezing height ($h$) while fixing the emission radius also yields similar results. Hence, we identify a region at $r_{em}\sim 5R$ where the PA becomes flat. This is shown by the blue dots obtained at $r_{em}=5R$ and $h=0.4R$ in panel (a) figure \ref{fig:PAcomp430}. 
{We see that for the higher intensity pulse, our model roughly reproduces the observed PA trend, but there is a significant discrepancy for the low intensity pulse.}

\begin{figure*}
    \includegraphics[scale=0.48]{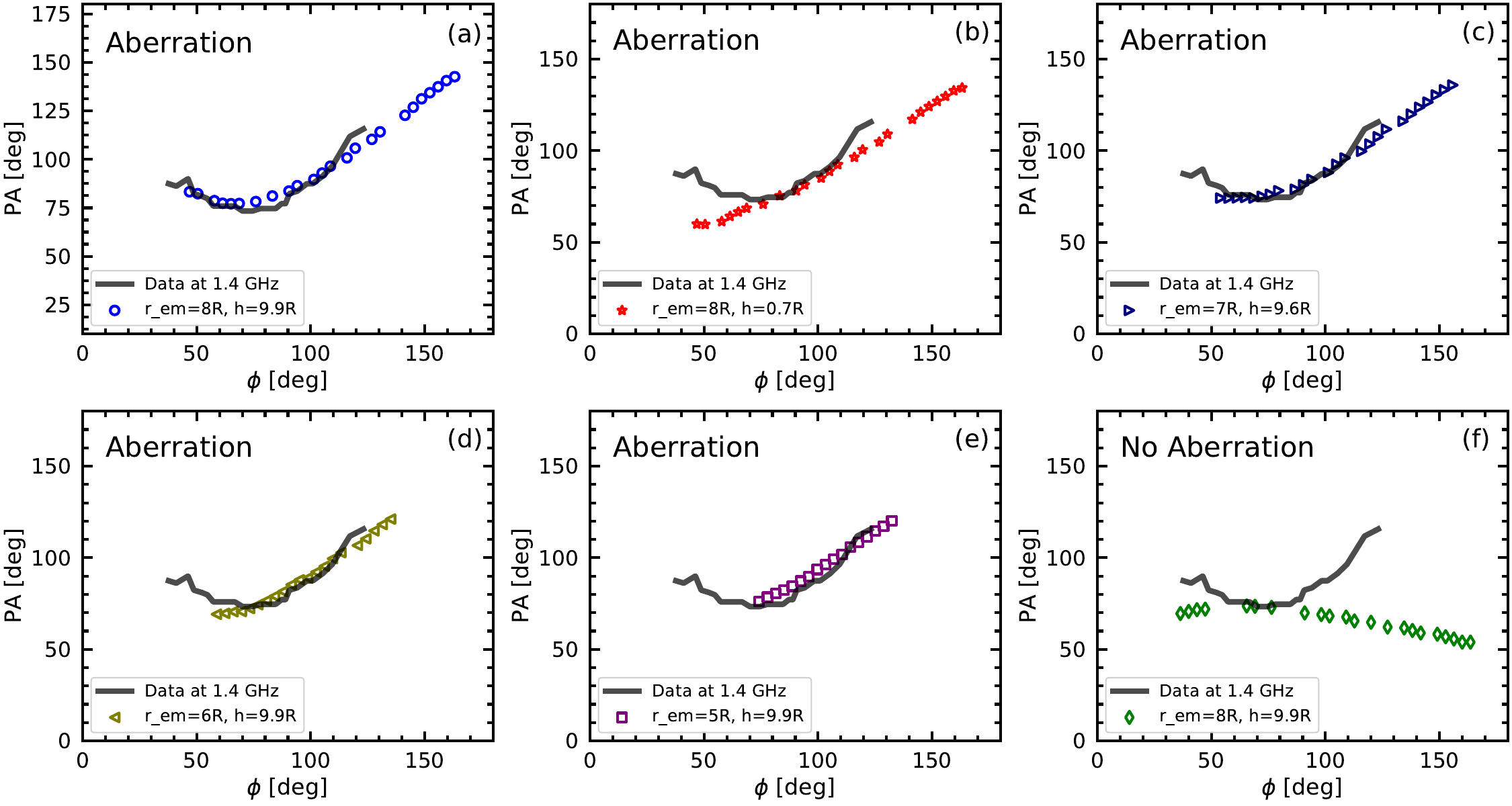}
    \caption{Comparison between our models and observed PA of PSR J0030+0451 at 1.4 GHz corresponding to the smaller intensity peak in its pulse profile.}
    \label{fig:PAcomp1400}
\end{figure*}
In either decreasing or increasing $h$, an abrupt discontinuity is seen at ${\phi\sim290^{\circ}}$, as shown by the squares obtained at $r_{em}=5R$ and $h=1.0R$ (panel b). {This feature is only present when aberration is taken into account in which case the $E_y$ component becomes close to zero along the trajectory.} Without aberration, the curve reverses its direction as shown by the green diamonds in panel (d). On the other hand, when neglecting the effect of aberration and choosing a higher emission radius/freezing height, for example $r_{em} = 7R$ and $h>4R$ (panel e), or $r_{em} = 8R$ (panel f), it appears that the observed PA trend for both pulses can be more or less reproduced.
However, the aberration effect is particularly important for MSPs and should not be neglected. Our failure to reproduce the PA trend for both pulse peaks at 430 MHz when aberration is included may suggest that either the other propagation effects need to be taken into account, like GR light bending close to the neutron star \citep{2020A&A...641A.166P}, and refraction of the radio waves by the plasma \citep{Beskin2012}, or the field configuration obtained by \cite{Chen2020} is not the actual field of PSR J0030+0451. A few other possible field geometries exist, as shown by \cite{Kalapotharakos2021}. Our modeling indicates that radio polarization is a more stringent test that may be able to break the degeneracy and select a more realistic field configuration.

\begin{figure}
    \includegraphics[scale=0.55]{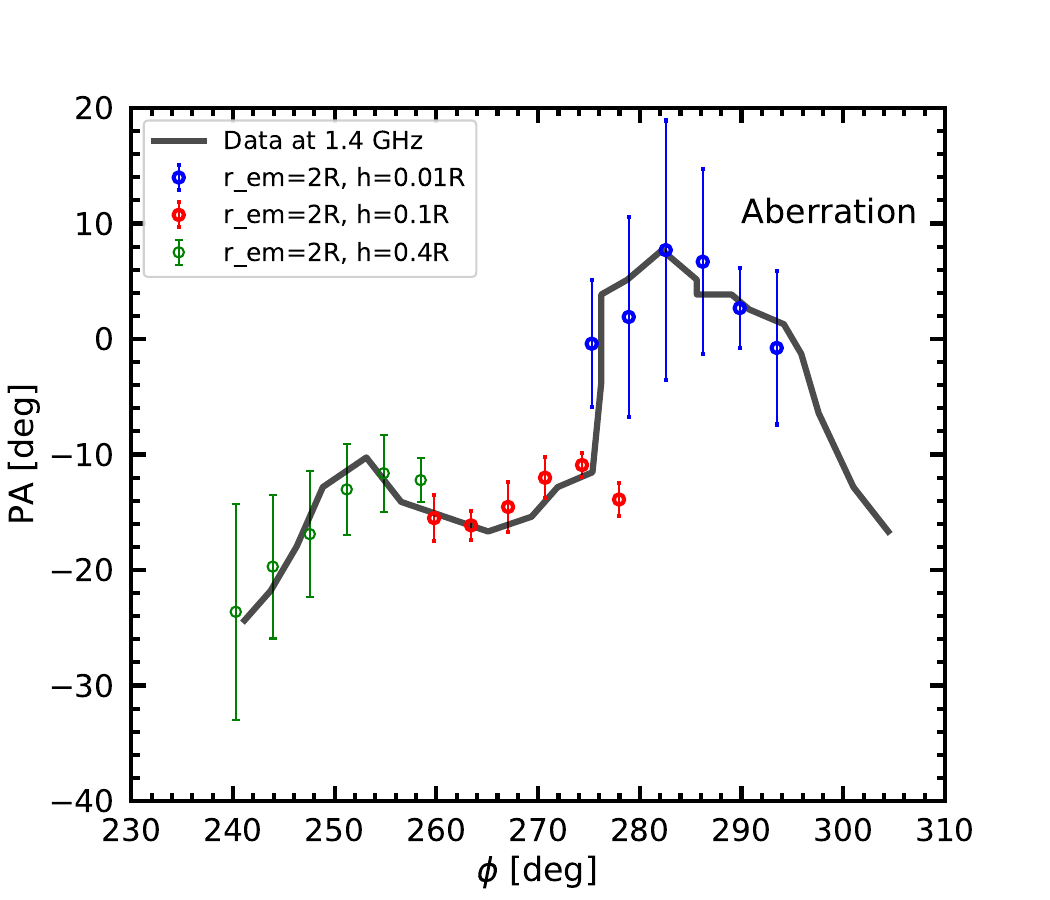}
    \caption{Comparing our models with the observed PA for the higher intensity pulse at 1.4 GHz with error bars corresponding to 100\% variation in the freezing height. }
    \label{fig:PAcomp1400b}
\end{figure}
Next, we examine the PA curve at 1.4 GHz, which again has two parts corresponding to two different intensity peaks in its pulse profile (see figure 1 in \citealt{Gentile2018}). The PA for the lower intensity pulse, shown in figure \ref{fig:PAcomp1400}, resembles the tail of an S-shaped curve resulting from a dipolar magnetic field. To explain this, we explore emission radii/freezing height greater than $5R$ because the dipolar components of the magnetic field are expected to become dominant over the quadrupolar components at larger distances. As we can see, the PA curve considering aberration and corresponding to $r_{em} = 8R$, and $h=9.9R$ exhibit a similar S-shaped swing (panel (a) in figure \ref{fig:PAcomp1400}). Decreasing the emission radius (panel e) or the freezing height (panel b) makes the PA curve straight. Furthermore, neglecting the effect of aberration with the same set of parameters causes the PA curve to be concave (panel f) and thus different from the observed feature. A comparison of panels (a), (c), and (d) in figure \ref{fig:PAcomp1400} indicates that a higher freezing height is necessary to recover the observed PA. Furthermore, we cannot go lower than $r_{em}=5R$ because the line of sight of the observer does not cross through this magnetic pole. \footnote{However, note that GR light bending and refraction of light can become important at small emission radii, and should be taken into account.} 

In order to explain the PA corresponding to the higher intensity peak which looks non-dipolar, we study emission coming from close to the surface of the NS. The PA is divided into three sections: an initial rise at the beginning, a distorted shape in the middle, and a decreasing tail at the end (figure \ref{fig:PAcomp1400b}). It turns out that this feature can be explained by varying the freezing height along the pulse. As the polarization is set when the emission decouples from the plasma, i.e. when the density sufficiently drops, decoupling {at different distances} might occur because of the plasma density distribution across magnetic field lines. The observed features can be recovered at $r_{em}=2R$, starting with $h=0.4R$ for the initial rise, $h=0.1R$ for the middle, and $h=0.01R$ for the decreasing tail. The error bars  (standard deviation) correspond to a variation of $100\%$ in $h$ and the PA(s) are reported as the mean value. {We note that very different emission radius $r_{\rm em}$ and freezing height $h$ are needed at the two poles in order to reproduce the observed PA trend at 1.4 GHz. Although the plasma density at the two poles could be different, leading to different $r_{\rm em}$ and $h$, it may be hard to account for such a large difference. This may again indicate that we either need to consider propagation effects, in particular GR light bending and plasma refraction close to the neutron star, or our field configuration is not the exact field possessed by PSR J0030+0451.}

\section{Conclusions and Discussions} \label{sec:conclusions}
Polarization modeling is a powerful tool and a very stringent test to constrain the magnetic field configuration of pulsars. An effective field model must pass all tests, including radio, X-ray, gamma-ray light curves, and radio polarization. In this paper, we compared radio polarization with observations for the MSP PSR J0030+0451 (which has detailed X-ray hotspot modeling) based on a multipolar magnetic field configuration, and we present easy equations for calculating the PA that can be used in parameter search models; if all physics is parameterized using $r_{em}$ and $h$.

We used the force-free configuration obtained by \cite{Chen2020} that can simultaneously produce the radio, X-ray, and gamma-ray light curves. In our models we calculate the polarization of plasma normal modes, taking into account the co-rotational motion of the plasma. We derive an expression of the PA by considering an O-mode instead of calculating the properties of the emission of charged particles in vacuum as described in \cite{BCW1991}. We modified the RVM to see if the multipolar field configuration could produce the observed radio polarization features. To determine the PA curves, we first created sky maps of the polarization taking into account aberration, and then selected a viewing angle. 
For PSR J0031+0451, the viewing angle was fixed at $\theta_{obs}=54^{\circ}$. We were able to reproduce some of the observed features of the PA swing both in the case of 430 MHz and 1.4 GHz observations by constraining the emission heights in each case at which emission is produced.

{When considering aberration, for 430 MHz, the PA swing at the higher intensity pulse could be explained by emissions coming from an effective radius $5R$ with $h=0.4R$.  However, for the lower-intensity pulse, the PA swing couldn't be reproduced at any chosen emission radius/height. On the other hand, without taking aberration into account, both pulses can be roughly reproduced simultaneously by emissions 
{from an effective radius greater than}
$7R$. It is not possible to explain simultaneously high and low intensity pulses at 1.4 GHz by any emission radius. Hence, we examined effective emission radii and heights separately for the two pulses to determine whether the observed features were reproducible. Our first finding was that when aberration was not considered, the PA swing for the smaller intensity pulse at any radius did not match the data. Therefore, we considered aberration in all our models for 1.4 GHz. Considering emissions 
{from a radius greater} 
than $8R$, it is possible to determine a similar PA swing for the smaller intensity pulse that resembles the tail of an S-curve. The higher intensity pulse with a distorted PA swing could be explained by emission from $2R$ and three different freezing heights. We note, therefore, that an important physics behind the MSP radio polarization model is including aberration effect, since without it many of the observed characteristics cannot be explained.}

All our models, however, ignore propagation effects. 
Radio waves propagate through dense magnetospheric plasma, where the polarization signature first evolves adiabatically following the magnetic field, before becoming permanent as they reach a larger distance where the plasma density has dropped \citep[e.g.,][]{Petrova2000,Wang2010,Beskin2012}. Effects like wave refraction, cyclotron absorption, and transition from geometrical optics to vacuum propagation can all influence the observed polarization. These propagation effects will depend on the properties of the magnetospheric plasma, e.g., the plasma density and its radial dependence, the drift motion of plasma particles, and the distribution function of the outgoing plasma. {It has also been found that pulsars whose PA profiles are not fitted with the RVM exhibit a much higher fraction of circular polarization than those with linear polarization \citep[e.g.,][]{Johnston2023}.} {Circular polarization is usually considered to be a result of propagation effects in the magnetospheric plasma \citep[e.g.,][]{Melrose2004}}. Furthermore, if the emission is produced close to the neutron star surface, GR light bending also needs to be taken into account \citep[e.g.,][]{Beloborodov2002, 2020A&A...641A.166P}, which we neglected in this work. A complete framework thus needs to include all the ingredients: a self-consistent magnetic field configuration, GR light bending, and propagation effects in the magnetospheric plasma. We plan to develop such a framework for polarization modeling in the future and combine it with multiwavelength light curve fitting. As a result of high-precision observational measurements, we may be able to constrain not only magnetic field configurations and radio emission sites but also magnetospheric plasma properties, which include density, bulk Lorentz factor, etc. This will give us further insights into the properties of pulsar plasma and radio emission physics. 

The multiwavelength light curve and radio polarization modeling of PSR J0030+0451 suggest that a multipolar magnetic field may be important near the neutron star surface. {A new analysis of {the NICER data for} PSR J0030+0451 {to constrain the size and location of its X-ray hot spots} shows that there is a multi-modal structure in the posterior surface {\citep{2023arXiv230809469V}. Given the uncertainty and degeneracy in X-ray fitting, radio polarization may turn out to be an important constraint to help distinguish between different field configurations.}} 
Obtaining a good understanding of the field configuration can have implications for other branches of pulsar physics as well, including the interior magnetic field evolution, formation of gravitational-wave mountains, and damping of r-modes when considering a superconducting core. These will be investigated in our future works.


\section*{Acknowledgements}
A.S gratefully acknowledges support and hospitality from the Simons Foundation through the predoctoral program at the Center for Computational Astrophysics, Flatiron Institute. Y.Y. would like to thank Alex Chen for helpful discussion. This work was also facilitated by Multimessenger Plasma Physics Center (MPPC), NSF grant PHY-2206608, and by a grant from the Simons Foundation (MP-SCMPS-00001470) to YY and AP.

\bibliography{sample631}{}

\begin{thebibliography}{}
\expandafter\ifx\csname natexlab\endcsname\relax\def\natexlab#1{#1}\fi
\providecommand{\url}[1]{\href{#1}{#1}}
\providecommand{\dodoi}[1]{doi:~\href{http://doi.org/#1}{\nolinkurl{#1}}}
\providecommand{\doeprint}[1]{\href{http://ascl.net/#1}{\nolinkurl{http://ascl.net/#1}}}
\providecommand{\doarXiv}[1]{\href{https://arxiv.org/abs/#1}{\nolinkurl{https://arxiv.org/abs/#1}}}

\bibitem[{{Alpar} {et~al.}(1982){Alpar}, {Cheng}, {Ruderman}, \& {Shaham}}]{Alpar1982}
{Alpar}, M.~A., {Cheng}, A.~F., {Ruderman}, M.~A., \& {Shaham}, J. 1982, \nat, 300, 728, \dodoi{10.1038/300728a0}

\bibitem[{{Arzoumanian} {et~al.}(2018){Arzoumanian}, {Baker}, {Brazier}, {Burke-Spolaor}, {Chamberlin}, {Chatterjee}, {Christy}, {Cordes}, {Cornish}, {Crawford}, {Thankful Cromartie}, {Crowter}, {DeCesar}, {Demorest}, {Dolch}, {Ellis}, {Ferdman}, {Ferrara}, {Folkner}, {Fonseca}, {Garver-Daniels}, {Gentile}, {Haas}, {Hazboun}, {Huerta}, {Islo}, {Jones}, {Jones}, {Kaplan}, {Kaspi}, {Lam}, {Lazio}, {Levin}, {Lommen}, {Lorimer}, {Luo}, {Lynch}, {Madison}, {McLaughlin}, {McWilliams}, {Mingarelli}, {Ng}, {Nice}, {Park}, {Pennucci}, {Pol}, {Ransom}, {Ray}, {Rasskazov}, {Siemens}, {Simon}, {Spiewak}, {Stairs}, {Stinebring}, {Stovall}, {Swiggum}, {Taylor}, {Vallisneri}, {van Haasteren}, {Vigeland}, {Zhu}, \& {NANOGrav Collaboration}}]{Nano2018}
{Arzoumanian}, Z., {Baker}, P.~T., {Brazier}, A., {et~al.} 2018, \apj, 859, 47, \dodoi{10.3847/1538-4357/aabd3b}

\bibitem[{{Arzoumanian} {et~al.}(2020){Arzoumanian}, {Baker}, {Blumer}, {B{\'e}csy}, {Brazier}, {Brook}, {Burke-Spolaor}, {Chatterjee}, {Chen}, {Cordes}, {Cornish}, {Crawford}, {Cromartie}, {Decesar}, {Demorest}, {Dolch}, {Ellis}, {Ferrara}, {Fiore}, {Fonseca}, {Garver-Daniels}, {Gentile}, {Good}, {Hazboun}, {Holgado}, {Islo}, {Jennings}, {Jones}, {Kaiser}, {Kaplan}, {Kelley}, {Key}, {Laal}, {Lam}, {Lazio}, {Lorimer}, {Luo}, {Lynch}, {Madison}, {McLaughlin}, {Mingarelli}, {Ng}, {Nice}, {Pennucci}, {Pol}, {Ransom}, {Ray}, {Shapiro-Albert}, {Siemens}, {Simon}, {Spiewak}, {Stairs}, {Stinebring}, {Stovall}, {Sun}, {Swiggum}, {Taylor}, {Turner}, {Vallisneri}, {Vigeland}, {Witt}, \& {Nanograv Collaboration}}]{Nano2020}
{Arzoumanian}, Z., {Baker}, P.~T., {Blumer}, H., {et~al.} 2020, \apjl, 905, L34, \dodoi{10.3847/2041-8213/abd401}

\bibitem[{{Backer} {et~al.}(1976){Backer}, {Rankin}, \& {Campbell}}]{Backer1976}
{Backer}, D.~C., {Rankin}, J.~M., \& {Campbell}, D.~B. 1976, \nat, 263, 202, \dodoi{10.1038/263202a0}

\bibitem[{{Bai} \& {Spitkovsky}(2010)}]{Bai2010}
{Bai}, X.-N., \& {Spitkovsky}, A. 2010, \apj, 715, 1282, \dodoi{10.1088/0004-637X/715/2/1282}

\bibitem[{{Bailes}(1989)}]{Bailes1989}
{Bailes}, M. 1989, \apj, 342, 917, \dodoi{10.1086/167647}

\bibitem[{{Beloborodov}(2002)}]{Beloborodov2002}
{Beloborodov}, A.~M. 2002, \apjl, 566, L85, \dodoi{10.1086/33951110.48550/arXiv.astro-ph/0201117}

\bibitem[{{Beskin} \& {Philippov}(2012)}]{Beskin2012}
{Beskin}, V.~S., \& {Philippov}, A.~A. 2012, \mnras, 425, 814, \dodoi{10.1111/j.1365-2966.2012.20988.x}

\bibitem[{{Bilous} {et~al.}(2019){Bilous}, {Watts}, {Harding}, {Riley}, {Arzoumanian}, {Bogdanov}, {Gendreau}, {Ray}, {Guillot}, {Ho}, \& {Chakrabarty}}]{Bilous2019}
{Bilous}, A.~V., {Watts}, A.~L., {Harding}, A.~K., {et~al.} 2019, \apjl, 887, L23, \dodoi{10.3847/2041-8213/ab53e7}

\bibitem[{{Blaskiewicz} {et~al.}(1991){Blaskiewicz}, {Cordes}, \& {Wasserman}}]{BCW1991}
{Blaskiewicz}, M., {Cordes}, J.~M., \& {Wasserman}, I. 1991, \apj, 370, 643, \dodoi{10.1086/169850}

\bibitem[{{Bogdanov} {et~al.}(2021){Bogdanov}, {Dittmann}, {Ho}, {Lamb}, {Mahmoodifar}, {Miller}, {Morsink}, {Riley}, {Strohmayer}, {Watts}, {Choudhury}, {Guillot}, {Harding}, {Ray}, {Wadiasingh}, {Wolff}, {Markwardt}, {Arzoumanian}, \& {Gendreau}}]{Bogdanov2021}
{Bogdanov}, S., {Dittmann}, A.~J., {Ho}, W. C.~G., {et~al.} 2021, \apjl, 914, L15, \dodoi{10.3847/2041-8213/abfb79}

\bibitem[{{Bransgrove} {et~al.}(2018){Bransgrove}, {Levin}, \& {Beloborodov}}]{Bransgrove2018}
{Bransgrove}, A., {Levin}, Y., \& {Beloborodov}, A. 2018, \mnras, 473, 2771, \dodoi{10.1093/mnras/stx2508}

\bibitem[{{Castillo} {et~al.}(2017){Castillo}, {Reisenegger}, \& {Valdivia}}]{Castillo2017}
{Castillo}, F., {Reisenegger}, A., \& {Valdivia}, J.~A. 2017, \mnras, 471, 507, \dodoi{10.1093/mnras/stx1604}

\bibitem[{{Chen} \& {Beloborodov}(2014)}]{Chen2014}
{Chen}, A.~Y., \& {Beloborodov}, A.~M. 2014, \apjl, 795, L22, \dodoi{10.1088/2041-8205/795/1/L22}

\bibitem[{{Chen} {et~al.}(2020){Chen}, {Yuan}, \& {Vasilopoulos}}]{Chen2020}
{Chen}, A.~Y., {Yuan}, Y., \& {Vasilopoulos}, G. 2020, \apjl, 893, L38, \dodoi{10.3847/2041-8213/ab85c5}

\bibitem[{{Chen} \& {Ruderman}(1993)}]{1993ApJ...408..179C}
{Chen}, K., \& {Ruderman}, M. 1993, \apj, 408, 179, \dodoi{10.1086/172578}

\bibitem[{{Chen} {et~al.}(1998){Chen}, {Ruderman}, \& {Zhu}}]{1998ApJ...493..397C}
{Chen}, K., {Ruderman}, M., \& {Zhu}, T. 1998, \apj, 493, 397, \dodoi{10.1086/305106}

\bibitem[{{Coles} {et~al.}(2015){Coles}, {Kerr}, {Shannon}, {Hobbs}, {Manchester}, {You}, {Bailes}, {Bhat}, {Burke-Spolaor}, {Dai}, {Keith}, {Levin}, {Os{\l}owski}, {Ravi}, {Reardon}, {Toomey}, {van Straten}, {Wang}, {Wen}, \& {Zhu}}]{Coles2015}
{Coles}, W.~A., {Kerr}, M., {Shannon}, R.~M., {et~al.} 2015, \apj, 808, 113, \dodoi{10.1088/0004-637X/808/2/113}

\bibitem[{{Cumming} {et~al.}(2004){Cumming}, {Arras}, \& {Zweibel}}]{Cumming2004}
{Cumming}, A., {Arras}, P., \& {Zweibel}, E. 2004, \apj, 609, 999, \dodoi{10.1086/421324}

\bibitem[{{Desvignes} {et~al.}(2019){Desvignes}, {Kramer}, {Lee}, {van Leeuwen}, {Stairs}, {Jessner}, {Cognard}, {Kasian}, {Lyne}, \& {Stappers}}]{Desvignes2019}
{Desvignes}, G., {Kramer}, M., {Lee}, K., {et~al.} 2019, Science, 365, 1013, \dodoi{10.1126/science.aav7272}

\bibitem[{{Dyks} {et~al.}(2004){Dyks}, {Harding}, \& {Rudak}}]{Dyks2004}
{Dyks}, J., {Harding}, A.~K., \& {Rudak}, B. 2004, \apj, 606, 1125, \dodoi{10.1086/383121}

\bibitem[{{Galishnikova} {et~al.}(2020){Galishnikova}, {Philippov}, \& {Beskin}}]{Alisa2020}
{Galishnikova}, A.~K., {Philippov}, A.~A., \& {Beskin}, V.~S. 2020, \mnras, 497, 2831, \dodoi{10.1093/mnras/staa2070}

\bibitem[{{Gentile} {et~al.}(2018){Gentile}, {McLaughlin}, {Demorest}, {Stairs}, {Arzoumanian}, {Crowter}, {Dolch}, {DeCesar}, {Ellis}, {Ferdman}, {Ferrara}, {Fonseca}, {Gonzalez}, {Jones}, {Jones}, {Lam}, {Levin}, {Lorimer}, {Lynch}, {Ng}, {Nice}, {Pennucci}, {Ransom}, {Ray}, {Spiewak}, {Stovall}, {Swiggum}, \& {Zhu}}]{Gentile2018}
{Gentile}, P.~A., {McLaughlin}, M.~A., {Demorest}, P.~B., {et~al.} 2018, \apj, 862, 47, \dodoi{10.3847/1538-4357/aac9c9}

\bibitem[{{Gil} {et~al.}(2006){Gil}, {Melikidze}, \& {Zhang}}]{Gil2006}
{Gil}, J., {Melikidze}, G., \& {Zhang}, B. 2006, \apj, 650, 1048, \dodoi{10.1086/506982}

\bibitem[{{Goldreich} \& {Reisenegger}(1992)}]{Goldreich1992}
{Goldreich}, P., \& {Reisenegger}, A. 1992, \apj, 395, 250, \dodoi{10.1086/171646}

\bibitem[{{Gourgouliatos} \& {Cumming}(2014)}]{Gourgouliatos2014}
{Gourgouliatos}, K.~N., \& {Cumming}, A. 2014, \mnras, 438, 1618, \dodoi{10.1093/mnras/stt2300}

\bibitem[{{Gupta} \& {Gangadhara}(2003)}]{Gupta2003}
{Gupta}, Y., \& {Gangadhara}, R.~T. 2003, \apj, 584, 418, \dodoi{10.1086/345682}

\bibitem[{{Hakobyan} {et~al.}(2019){Hakobyan}, {Philippov}, \& {Spitkovsky}}]{Hayk2019}
{Hakobyan}, H., {Philippov}, A., \& {Spitkovsky}, A. 2019, \apj, 877, 53, \dodoi{10.3847/1538-4357/ab191b}

\bibitem[{{Hakobyan} {et~al.}(2017){Hakobyan}, {Beskin}, \& {Philippov}}]{Hakobyan2017}
{Hakobyan}, H.~L., {Beskin}, V.~S., \& {Philippov}, A.~A. 2017, \mnras, 469, 2704, \dodoi{10.1093/mnras/stx1025}

\bibitem[{{Johnston} {et~al.}(2023){Johnston}, {Kramer}, {Karastergiou}, {Keith}, {Oswald}, {Parthasarathy}, \& {Weltevrede}}]{Johnston2023}
{Johnston}, S., {Kramer}, M., {Karastergiou}, A., {et~al.} 2023, \mnras, 520, 4801, \dodoi{10.1093/mnras/stac3636}

\bibitem[{{Jones} {et~al.}(2017){Jones}, {McLaughlin}, {Lam}, {Cordes}, {Levin}, {Chatterjee}, {Arzoumanian}, {Crowter}, {Demorest}, {Dolch}, {Ellis}, {Ferdman}, {Fonseca}, {Gonzalez}, {Jones}, {Lazio}, {Nice}, {Pennucci}, {Ransom}, {Stinebring}, {Stairs}, {Stovall}, {Swiggum}, \& {Zhu}}]{Jones2017}
{Jones}, M.~L., {McLaughlin}, M.~A., {Lam}, M.~T., {et~al.} 2017, \apj, 841, 125, \dodoi{10.3847/1538-4357/aa73df}

\bibitem[{{Kalapotharakos} \& {Contopoulos}(2009)}]{Kalapotharakos2009}
{Kalapotharakos}, C., \& {Contopoulos}, I. 2009, \aap, 496, 495, \dodoi{10.1051/0004-6361:200810281}

\bibitem[{{Kalapotharakos} {et~al.}(2021){Kalapotharakos}, {Wadiasingh}, {Harding}, \& {Kazanas}}]{Kalapotharakos2021}
{Kalapotharakos}, C., {Wadiasingh}, Z., {Harding}, A.~K., \& {Kazanas}, D. 2021, \apj, 907, 63, \dodoi{10.3847/1538-4357/abcec0}

\bibitem[{{Kramer} \& {Wex}(2009)}]{Kramer2009}
{Kramer}, M., \& {Wex}, N. 2009, Classical and Quantum Gravity, 26, 073001, \dodoi{10.1088/0264-9381/26/7/073001}

\bibitem[{{Kramer} {et~al.}(1998){Kramer}, {Xilouris}, {Lorimer}, {Doroshenko}, {Jessner}, {Wielebinski}, {Wolszczan}, \& {Camilo}}]{Kramer1998}
{Kramer}, M., {Xilouris}, K.~M., {Lorimer}, D.~R., {et~al.} 1998, \apj, 501, 270, \dodoi{10.1086/305790}

\bibitem[{{Kramer} {et~al.}(2006){Kramer}, {Stairs}, {Manchester}, {McLaughlin}, {Lyne}, {Ferdman}, {Burgay}, {Lorimer}, {Possenti}, {D'Amico}, {Sarkissian}, {Hobbs}, {Reynolds}, {Freire}, \& {Camilo}}]{Kramer2006}
{Kramer}, M., {Stairs}, I.~H., {Manchester}, R.~N., {et~al.} 2006, Science, 314, 97, \dodoi{10.1126/science.1132305}

\bibitem[{{Lorimer}(2008)}]{Lorimer2008}
{Lorimer}, D.~R. 2008, Living Reviews in Relativity, 11, 8, \dodoi{10.12942/lrr-2008-8}

\bibitem[{{Manchester} \& {Taylor}(1977)}]{Manchester1977}
{Manchester}, R.~N., \& {Taylor}, J.~H. 1977, {Pulsars} (San Francisco : W. H. Freeman)

\bibitem[{{Manchester} {et~al.}(1975){Manchester}, {Taylor}, \& {Huguenin}}]{Manchester1975}
{Manchester}, R.~N., {Taylor}, J.~H., \& {Huguenin}, G.~R. 1975, \apj, 196, 83, \dodoi{10.1086/153395}

\bibitem[{{Melatos} \& {Phinney}(2001)}]{2001PASA...18..421M}
{Melatos}, A., \& {Phinney}, E.~S. 2001, \pasa, 18, 421, \dodoi{10.1071/AS01056}

\bibitem[{{Melrose} \& {Luo}(2004)}]{Melrose2004}
{Melrose}, D.~B., \& {Luo}, Q. 2004, \mnras, 352, 915, \dodoi{10.1111/j.1365-2966.2004.07986.x}

\bibitem[{{Melrose} {et~al.}(2021){Melrose}, {Rafat}, \& {Mastrano}}]{Melrose2021}
{Melrose}, D.~B., {Rafat}, M.~Z., \& {Mastrano}, A. 2021, \mnras, 500, 4549, \dodoi{10.1093/mnras/staa3529}

\bibitem[{{Miller} {et~al.}(2019){Miller}, {Lamb}, {Dittmann}, {Bogdanov}, {Arzoumanian}, {Gendreau}, {Guillot}, {Harding}, {Ho}, {Lattimer}, {Ludlam}, {Mahmoodifar}, {Morsink}, {Ray}, {Strohmayer}, {Wood}, {Enoto}, {Foster}, {Okajima}, {Prigozhin}, \& {Soong}}]{Miller2019}
{Miller}, M.~C., {Lamb}, F.~K., {Dittmann}, A.~J., {et~al.} 2019, \apjl, 887, L24, \dodoi{10.3847/2041-8213/ab50c5}

\bibitem[{{Mitra} {et~al.}(2023){Mitra}, {Melikidze}, \& {Basu}}]{Mitra2023}
{Mitra}, D., {Melikidze}, G.~I., \& {Basu}, R. 2023, \apj, 952, 151, \dodoi{10.3847/1538-4357/acdc28}

\bibitem[{{Payne} \& {Melatos}(2004)}]{2004MNRAS.351..569P}
{Payne}, D.~J.~B., \& {Melatos}, A. 2004, \mnras, 351, 569, \dodoi{10.1111/j.1365-2966.2004.07798.x}

\bibitem[{{P{\'e}tri}(2019)}]{Petri2019}
{P{\'e}tri}, J. 2019, \mnras, 484, 5669, \dodoi{10.1093/mnras/stz360}

\bibitem[{{Petrova} \& {Lyubarskii}(2000)}]{Petrova2000}
{Petrova}, S.~A., \& {Lyubarskii}, Y.~E. 2000, \aap, 355, 1168

\bibitem[{{Philippov} \& {Kramer}(2022)}]{Philippov2022}
{Philippov}, A., \& {Kramer}, M. 2022, \araa, 60, 495, \dodoi{10.1146/annurev-astro-052920-112338}

\bibitem[{{Philippov} {et~al.}(2020){Philippov}, {Timokhin}, \& {Spitkovsky}}]{Philippov2020}
{Philippov}, A., {Timokhin}, A., \& {Spitkovsky}, A. 2020, \prl, 124, 245101, \dodoi{10.1103/PhysRevLett.124.245101}

\bibitem[{{Philippov} \& {Spitkovsky}(2018)}]{Philippov2018}
{Philippov}, A.~A., \& {Spitkovsky}, A. 2018, \apj, 855, 94, \dodoi{10.3847/1538-4357/aaabbc}

\bibitem[{{Pons} \& {Geppert}(2007)}]{Pons2007}
{Pons}, J.~A., \& {Geppert}, U. 2007, \aap, 470, 303, \dodoi{10.1051/0004-6361:20077456}

\bibitem[{{Poutanen}(2020)}]{2020A&A...641A.166P}
{Poutanen}, J. 2020, \aap, 641, A166, \dodoi{10.1051/0004-6361/202038689}

\bibitem[{{Raaijmakers} {et~al.}(2019){Raaijmakers}, {Riley}, {Watts}, {Greif}, {Morsink}, {Hebeler}, {Schwenk}, {Hinderer}, {Nissanke}, {Guillot}, {Arzoumanian}, {Bogdanov}, {Chakrabarty}, {Gendreau}, {Ho}, {Lattimer}, {Ludlam}, \& {Wolff}}]{Raajimakers2019}
{Raaijmakers}, G., {Riley}, T.~E., {Watts}, A.~L., {et~al.} 2019, \apjl, 887, L22, \dodoi{10.3847/2041-8213/ab451a}

\bibitem[{{Radhakrishnan}(1984)}]{Radhakrishnan1984}
{Radhakrishnan}, V. 1984, in Birth and Evolution of Neutron Stars: Issues Raised by Millisecond Pulsars, ed. S.~P. {Reynolds} \& D.~R. {Stinebring}, 130

\bibitem[{{Radhakrishnan} \& {Cooke}(1969)}]{Radhakrishnan1969}
{Radhakrishnan}, V., \& {Cooke}, D.~J. 1969, \aplett, 3, 225

\bibitem[{{Rankin} {et~al.}(2017){Rankin}, {Archibald}, {Hessels}, {van Leeuwen}, {Mitra}, {Ransom}, {Stairs}, {van Straten}, \& {Weisberg}}]{Rankin2017}
{Rankin}, J.~M., {Archibald}, A., {Hessels}, J., {et~al.} 2017, \apj, 845, 23, \dodoi{10.3847/1538-4357/aa7b73}

\bibitem[{{Riley} {et~al.}(2019){Riley}, {Watts}, {Bogdanov}, {Ray}, {Ludlam}, {Guillot}, {Arzoumanian}, {Baker}, {Bilous}, {Chakrabarty}, {Gendreau}, {Harding}, {Ho}, {Lattimer}, {Morsink}, \& {Strohmayer}}]{Riley2019}
{Riley}, T.~E., {Watts}, A.~L., {Bogdanov}, S., {et~al.} 2019, \apjl, 887, L21, \dodoi{10.3847/2041-8213/ab481c}

\bibitem[{{Romani}(1990)}]{1990Natur.347..741R}
{Romani}, R.~W. 1990, \nat, 347, 741, \dodoi{10.1038/347741a0}

\bibitem[{{Ruderman}(1991)}]{1991ApJ...366..261R}
{Ruderman}, M. 1991, \apj, 366, 261, \dodoi{10.1086/169558}

\bibitem[{{Spitkovsky}(2006)}]{Spitkovsky2006}
{Spitkovsky}, A. 2006, \apjl, 648, L51, \dodoi{10.1086/507518}

\bibitem[{{Sur} {et~al.}(2020){Sur}, {Haskell}, \& {Kuhn}}]{Sur2020}
{Sur}, A., {Haskell}, B., \& {Kuhn}, E. 2020, \mnras, 495, 1360, \dodoi{10.1093/mnras/staa1212}

\bibitem[{{Vinciguerra} {et~al.}(2023){Vinciguerra}, {Salmi}, {Watts}, {Choudhury}, {Riley}, {Ray}, {Bogdanov}, {Kini}, {Guillot}, {Chakrabarty}, {Ho}, {Huppenkothen}, {Morsink}, \& {Wadiasingh}}]{2023arXiv230809469V}
{Vinciguerra}, S., {Salmi}, T., {Watts}, A.~L., {et~al.} 2023, arXiv e-prints, arXiv:2308.09469, \dodoi{10.48550/arXiv.2308.09469}

\bibitem[{Wang \& Lai(2007)}]{Wang2007}
Wang, C., \& Lai, D. 2007, Monthly Notices of the Royal Astronomical Society, 377, 1095, \dodoi{10.1111/j.1365-2966.2007.11531.x}

\bibitem[{{Wang} {et~al.}(2010){Wang}, {Lai}, \& {Han}}]{Wang2010}
{Wang}, C., {Lai}, D., \& {Han}, J. 2010, \mnras, 403, 569, \dodoi{10.1111/j.1365-2966.2009.16074.x}

\end{thebibliography}
\bibliographystyle{aasjournal}



\end{document}